\documentclass[preprint,prc,aps,showpacs]{revtex4-1}
\usepackage{graphicx}

\usepackage{amsmath} 
\usepackage{amssymb} 

\newcommand{\bra}[1]{\langle #1|}
\newcommand{\ket}[1]{|#1\rangle}
\newcommand{\braket}[2]{\langle #1|#2\rangle}

\begin{document}
\title{Obtaining the strong coupling constants $g_{J/\psi D_s D_s}$ and $g_{\phi D_s D_s }$  from QCD Sum Rules}
\author{M. E. Bracco}\email{bracco@uerj.br}
\affiliation{Faculdade de Tecnologia, Universidade do Estado do Rio de Janeiro, 
Rod. Presidente Dutra Km 298, P\'olo Industrial, 27537-000, Resende, RJ, Brazil.}

\author{M. Chiapparini and B. Os\'orio Rodrigues}
\affiliation{Instituto de F\'{\i}sica, Universidade do Estado do Rio de 
Janeiro, Rua S\~ao Francisco Xavier 524, 20550-900, Rio de Janeiro, RJ, Brazil. }

\begin{abstract}
The form factors and coupling constants of the meson vertices $J/\psi D_s D_s$ and $\phi D_s D_s$ were calculated using three point correlation functions within the QCD Sum Rules formalism. We have considered the cases where $\phi$, $D_s$ and $J/\psi$ mesons are off-shell obtaining, for each vertex, two different form factors and its corresponding coupling constants, namely $g_{J/\psi D_s D_s} = 6.20^{+0.97}_{-1.15}$ and $g_{\phi D_s D_s} = 1.85^{+0.22}_{-0.23}$.
\end{abstract}


\maketitle

\section{Introduction}
In the last decade, the increasing energies of the collision experiments have leaded to the detection of various new and ``exotic" mesons as, for example, the $Y(4140)$ (reported by CDF Collaboration as a narrow near-threshold structure in the $J/\psi\phi$ mass spectrum \cite{Aaltonen:2009}) and the $X(4350)$ (reported by Belle Collaboration also as another narrow structure in the $J/\psi\phi$ mass spectrum at $4.35$ GeV \cite{Shen:2010}). By exotic, we mean that it is believed that these mesons are beyond the usual quark-model description as  $q\bar{q}$ pairs. Since most of their quantum numbers have not been measured yet, there are many uncertainties about their constitution. Various models started to appear attempting to explain the masses and the observed decay modes of these mesons. Some of these models consider these exotic mesons as quark-gluon hybrids $(q\bar qg)$, tetraquark states $(q\bar qq\bar q)$, molecular states of two ordinary mesons, glueballs, states with exotic quantum numbers and many others \cite{Mahajan:2009pj,Branz:2009yt,Liu:2009iw,Liu:2009ei,Zhao:2011sd}. Some of these hypothesis have been studied in many works using the QCD Sum Rules (QCDSR) approach regarding masses, branching ratios and decay constants, obtaining different degrees of success depending on the choice of the interpolating currents \cite{Nielsen:2009uh, Albuquerque:2009ak,Zanetti:2011ju,Finazzo:2011he,Albuquerque:2010fm}.

Following the development of previous works of our group (\cite{Bracco:2004rx,Rodrigues:2010ed,Bracco:2011pg} and references therein, just to name a few), in this work we use the QCDSR formalism to obtain the coupling constants of the meson vertices $J/\psi D_s D_s$ and $\phi D_s D_s$. These vertices may appear in the decay processes $Y(4140) \to J/\psi \phi$ and  $X(4350) \to J/\psi \phi$, depending on the model used to represent the $Y(4140)$ and $X(4350)$ mesons. We call the attention to the presence of these two vertices in the works of Ref. \cite{Liu:2009iw}, where the observed decay $Y(4140) \to J/\psi \phi$ is studied  with the intermediate decay $Y(4140) \to D_s \bar{D}^{(*)}_s \to J/\psi \phi$, and Ref. \cite{Zhao:2011sd}, where the observed decay $X(4350) \to J/\psi \phi$ is analyzed using the intermediate process $X(4350) \to D_s^{(*)} \bar{D}_s^{(*)}  \to J/\psi \phi$. It is worth to mention that a more precise knowledge  of these coupling constants and form factors certainly will improve the understanding of the fundamental constitution of these new observed mesons.

\section{Formalism}
The starting point for a three meson vertex in QCDSR is the three point correlation function \cite{Bracco:2011pg}. In this work, we are interested in two vertices of the type vector-pseudoscalar-pseudoscalar ($VPP$), namely the $J/\psi D_sD_s$ and $\phi D_sD_s$ vertices. Therefore, we have two different correlation functions for each vertex, one with a vector meson off-shell and another with the pseudo-scalar meson ($D_s$) off-shell:
\begin{eqnarray}
\Gamma^{(V_n)}_\mu(p,p') &=& \int \bra{0'} T\{ j^{D_s}_5(x)j^{V_n\dagger}_\mu(y)j^{D_s\dagger}_5(0) \}\ket{0'} e^{ip'x}e^{-iqy}d^4x d^4y  \label{eq:pivoff} \\
\Gamma^{(D_s)}_\mu(p,p') &= &\int \bra{0'} T\{ j^{D_s}_5(x)j^{D_s\dagger}_5(y)j^{V_n\dagger}_\mu(0) \}\ket{0'} e^{ip'x}e^{-iqy}d^4x d^4y\,\;\;\;\;n=1,2 \label{eq:pidsoff}
\end{eqnarray}
where $V_1 = \phi$ and $V_2 = J/\psi$, being $q = p' - p$ the transferred momentum.

According to the QCDSR, we can calculate these correlation functions in two different ways: with hadron degrees of freedom (called the \textit{phenomenological side}) or with quark degrees of freedom (called the \textit{OPE side}). Both representations are equivalent and can be equated invoking the quark-hadron duality, after the application of a double Borel transform on each side. Then, the form factors and coupling constants can be extracted.

\subsection{The phenomenological side}
In order to accomplish the calculation of the phenomenological side, it is necessary to know the effective Lagrangians of the interaction which, for the vertices $V_n D_s D_s$, are \cite{Liu:2009iw,Zhao:2011sd}: 
\begin{align}
\mathcal{L}_{V_n D_s D_s} = (-1)^n i g_{V_n D_s D_s} V^\alpha_n ( D^+_s\partial_\alpha D^-_s - \partial_\alpha D^+_s D_s^-),\;\;\;\;n=1,2
\label{eq:lagrangeana}
\end{align}

From these Lagrangians, we can obtain the vertex of the hadronic process. In the case of  $V_n$ off-shell we have:
\begin{eqnarray}
\braket{D_s(p)V_n(q)}{D_s(p')} &=& (-1)^n i g^{(V_n)}_{V_n D_s D_s}(q^2)\epsilon^\alpha(q)(p_\alpha + p'_\alpha) 
\end{eqnarray}
We also make use of the hadronic matrix elements
\begin{eqnarray}
\bra{0} j_5^{D_s} \ket{D_s(p')} &=& \bra{D_s(p)} j_5^{D_s} \ket{0} = f_{D_s} \frac{m^2_{D_s}}{m_c + m_s}\\
\bra{V_n(q)} j_\mu^{V_n} \ket{0} &=& f_{V_n} m_{V_n} \epsilon^*_\mu(q)
\end{eqnarray}
In the case of  $D_s$ off-shell the obtained vertex is: 
\begin{eqnarray}
\braket{D_s(q)V_n(p)}{D_s(p')} &=& (-1)^n i g^{(D_s)}_{V_n D_s D_s}(q^2)\epsilon^\alpha(p)(2p'_\alpha - p_\alpha)
\end{eqnarray}
and the corresponding usual matrix elements are:
\begin{eqnarray}
\bra{0} j_5^{D_s} \ket{D_s(p')} &=& \bra{D_s(q)} j_5^{D_s} \ket{0} = f_{D_s} \frac{m^2_{D_s}}{m_c + m_s}\\
\bra{V_n(p)} j_\mu^{V_n} \ket{0} &=& f_{V_n} m_{V_n} \epsilon^*_\mu(p)
\end{eqnarray}
In the formulas above, $f_{V_n}$ and $f_{D_s}$ are the decay constants of the vector ($n=1,2$) and $D_s$ mesons respectively,  $\epsilon_{\mu(\alpha)}$ is the polarization vector, $g^{(V_n)}_{V_n D_s D_s}(q^2)$ and $g^{(D_s)}_{V_n D_s D_s}(q^2)$ are the form factors of the vertices, which correspond
to $V_n$ or $D_s$ messon off-shell respectivelly,  $m_s$, $m_c$ are the strange and 
charm quark masses and $m_{V_n}$  is the mass of the vector meson.

In order to improve the matching between the phenomenological and QCD sides, it is convenient to make the change of variables $p^2 \rightarrow -P^2$,  $p'^2 \rightarrow -{P'}^2$ and $q^2 \rightarrow -Q^2$. After that, we can obtain the correlation functions for the phenomenological side in the form:
\begin{eqnarray}
\Gamma^{phen (V_n)}_\mu &=& \frac{(-1)^n g^{(V_n)}_{V_n D_s D_s}(Q^2) f_{D_s}^2 f_{V_n} m^4_{D_s} m_{V_n}(p_\mu + p'_\mu)}{(m_c + m_s)^2 (P^2 + m_{D_s}^2)(Q^2 + m_{V_n}^2)(P'^2 + m_{D_s}^2)  } + h. r. \label{eq:fenomvnoff} \\
\Gamma^{phen (D_s)}_\mu &=& -\frac{(-1)^n g^{(D_s)}_{V_n D_s D_s}(Q^2) f_{D_s}^2 f_{V_n} m^4_{D_s} m_{V_n}\left ( \frac{Q^2 - P^2 - P'^2}{m^2_{V_n}} p_\mu - 2 p'_\mu \right )}{(m_c + m_s)^2 (P^2 + m_{V_n}^2)(Q^2 + m_{D_s}^2)(P'^2 + m_{D_s}^2)  } + h. r. \label{eq:fenomdsoff}
\end{eqnarray}
where $h.r.$ stands for the contributions of higher resonances and continuum states of the involved mesons and $g^{(M)}_{V_n D_s D_s}(Q^2)$ is the form factor of the $V_n D_s D_s$ vertex with meson $M$ ($M=V_n, D_s$) off-shell. We remind that the limit  $\lim\limits_{Q^2 \to -m_M^2} g^{(M)}_{V_n D_s D_s}(Q^2)$ is our definition for the coupling constant $g_{V_n D_s D_s}$ of the vertex \cite{Bracco:2001pqq}.

\subsection{The OPE side}
The OPE side is obtained using the interpolating currents written in terms of the quark fields in Eqs.~(\ref{eq:pivoff}) and (\ref{eq:pidsoff}). The meson interpolating currents used in this work are $j_\mu^{J/\psi} = \bar{c} \gamma_\mu c$, $j_\mu^{\phi} = \bar{s} \gamma_\mu s$ and $j_5^{D_s^-} = i \bar{c} \gamma_5 s$. 

By construction, the OPE side is given by an expansion known as Wilson's Operator Product Expansion. In the form factor calculation, this expansion exhibits a rapid convergence and can be truncated after a few terms. In this work, we consider contributions from two kinds of terms in the OPE side: the perturbative (the dominant one) and the quark condensates. In previous works of our group \cite{Bracco:2011pg,Matheus:2005yu}, it was shown that the gluon condensate term has a negligible contribution. For this reason, it has not been included in the present work. By using dispersion relations, the correlation function for a given meson $M$ off-shell can be written in the following form:
\begin{align}
\Gamma^{OPE (M)}_\mu(p,p') = - \frac{1}{4\pi^2} \int^\infty_0 \int^\infty_0 \frac{\rho_\mu^{pert (M)}(s, u, t)}{(s-p^2)(u-p'^2)} ds du + \Gamma^{\langle\bar{q}q\rangle (M)}_\mu
\label{eq:piladodaqcdgeral}
\end{align}
where the first and second terms are the perturbative and quark-condensate contributions respectively. The spectral density $\rho_\mu^{pert (M)}(s, u, t)$ is the double discontinuity of the perturbative term of the OPE, obtained by the use of the Cutkosky's rules and $s=p^2$, $u=p'^2$, and $t=q^2$ are the Mandelstam variables of the vertex.

Invoking Lorentz symmetry,  the spectral density for a $VPP$ vertex can be parametrized as:
\begin{align}
\rho_\mu^{pert (M)}(s,u,t) = \frac{3}{2\sqrt{\lambda}}\left [ F_p^{(M)}(s,u,t)p_\mu +  F_{p'}^{(M)}(s,u,t)p'_\mu \right ]
\label{dspec}
\end{align}
where $\lambda = (u + s - t)^2 - 4us$, and $F_p^{(M)}$ and $F_{p'}^{(M)}$ are invariant amplitudes. For the cases studied in this work, these invariant amplitudes can be written as:
\begin{eqnarray}
F_p^{(V_n)} &=& (-1)^n[u(A - 1) - A(t-s) - (2A - 1)(m_s-m_c)^2]\\
F_{p'}^{(V_n)} &=& (-1)^n[s(B - 1) + B(u - t) + (1 - 2B)(m_s-m_c)^2]\\
F_p^{(D_s)} &=& u(A + (-1)^n) + A(t-s) - ((-1)^n + 2A)(m_s-m_c)^2 \\
F_{p'}^{(D_s)} &=& -s(B + (-1)^n) + B(u + t) - 2B(m_s-m_c)^2
\end{eqnarray}
where 
\begin{align}
&A = \left [ \frac{\bar{k}_0}{\sqrt{s}} - \frac{p'_0 \overline{|\vec{k}|} \overline{\cos\theta}}{|\vec{p'}|\sqrt{s}} \right ]
\;\;\;\;\;\;\;\;
B = \frac{\overline{|\vec{k}|} \overline{\cos\theta}}{|\vec{p'}|} 
\;\;\;\;\;\;\;\;
\overline{\cos\theta} = \frac{2p'_0\bar{k}_0 - u + (-1)^{(n+\epsilon)}(m_s^2 - m_c^2) }{2|\vec{p'}|\overline{|\vec{k}|}} 
\nonumber \\
&\overline{|\vec{k}|} = \sqrt{\bar{k}_0^2 - |2-n-\epsilon|m_s^2 - |1-n+\epsilon|m_c^2}
\;\;\;\;\;\;\;\;
\bar{k}_0 = \frac{s + \epsilon(-1)^{(n+\epsilon)}(m_c^2 - m_s^2)}{2\sqrt{s}}
\nonumber \\
&p'_0 = \frac{s+u-t}{2\sqrt{s}}\;\;\;\;\;\;\;\;|\vec{p'}| = \frac{\sqrt{\lambda}}{2\sqrt{s}} \nonumber
\end{align}
and $\epsilon = 0(1)$ for $D_s(V_n)$ off-shell.

\begin{figure}[ht]
  \includegraphics[width=305px]{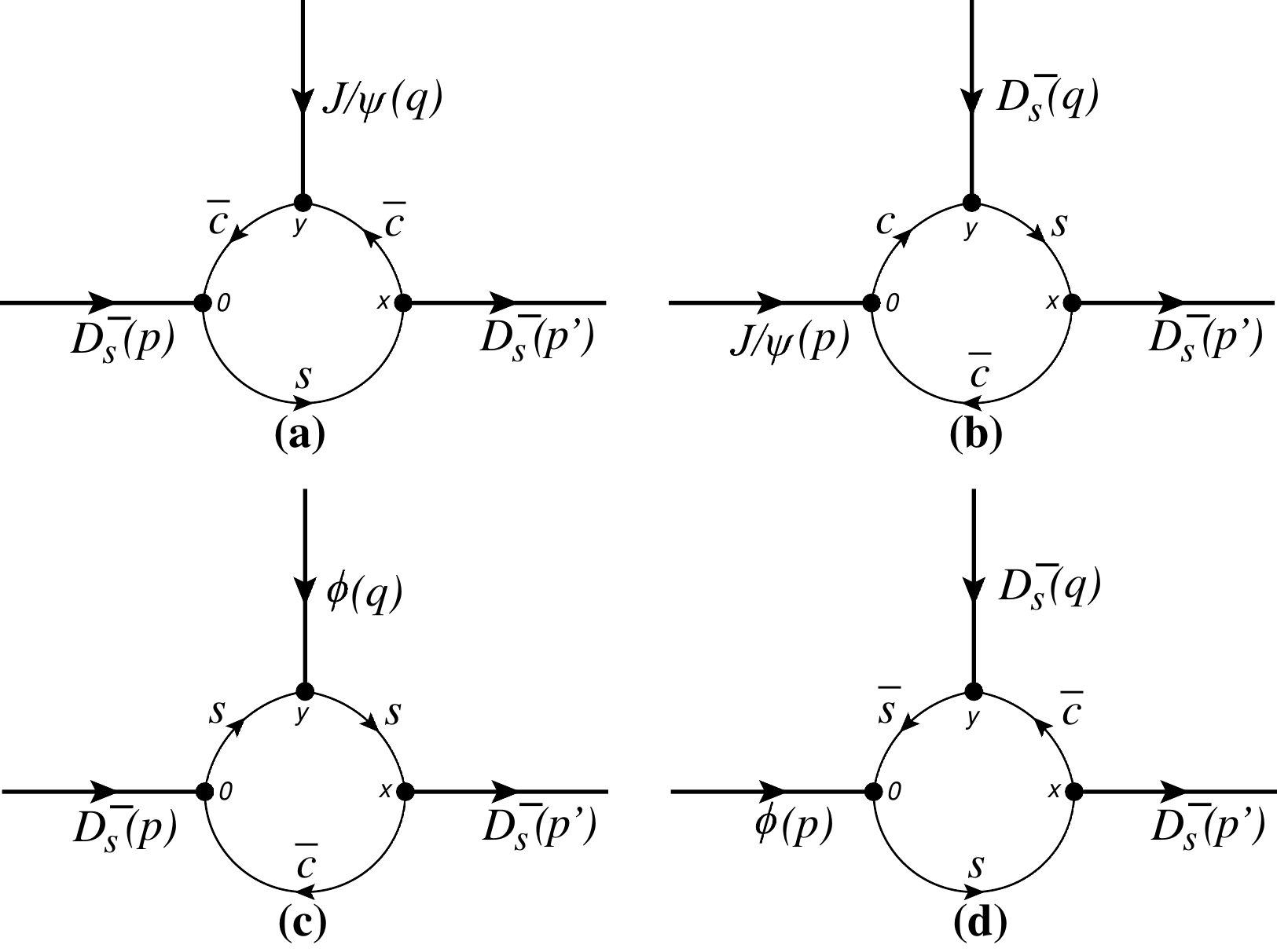}
  \caption{  \label{fig:percontrib}Perturbative contributions to vertex $J/\psi D_s D_s$ with $J/\psi$ off-shell (panel \textbf{(a)}) and $D_s$ off-shell (panel \textbf{(b)}), and to vertex $\phi D_s D_s$ with $\phi$ off-shell (panel \textbf{(c)}) and $D_s$ off-shell (panel \textbf{(d)}).}
\end{figure}

\begin{figure}[ht]
  \includegraphics[width=305px]{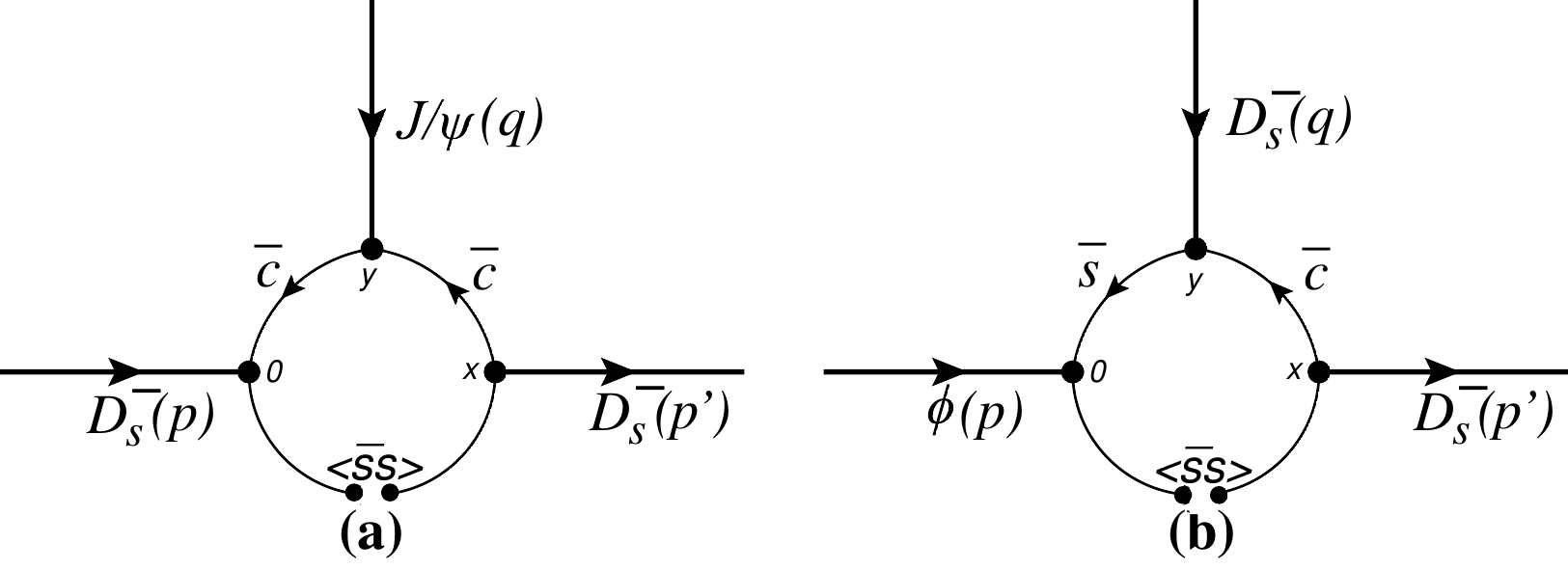}
  \caption{\label{fig:contribcond}Quark condensates contributing to vertices $J/\psi D_s D_s$ (panel \textbf{(a)}) and $\phi D_s D_s$ (panel \textbf{(b)}).}
  \end{figure}

Diagrammatically, the perturbative contributions are shown in Fig.~\ref{fig:percontrib}. The quark condensates contributing to Eq.~(\ref{eq:piladodaqcdgeral}) are: 
\begin{align}
\Gamma^{\langle\bar{s}s\rangle (M)}_\mu = -|n+\epsilon - 2|\langle\bar{s}s\rangle\frac{[m_c p_\mu + ((\epsilon - 1)m_s + \epsilon m_c)p'_\mu ]}{(p^2 - (1-\epsilon)m_s^2 - \epsilon m^2_c)(p'^2 - m_c^2)}
\label{eq:contribcond}
\end{align}
where $\langle\bar{s}s\rangle = (0.8 \pm 0.2)\langle\bar{q}q\rangle = (0.8 \pm 0.2)(0.23 \pm 0.03)^3$ GeV$^3$ \cite{Nielsen:2009uh}. The quark condensate contributions are represented by the diagrams of Fig.~\ref{fig:contribcond}.\\

\subsection{The sum rules}
To obtain the sum rules, two more steps are necessary. The first one is to apply a double Borel transform \cite{Khodjamirian:2002pka,Colangelo:2000dp} to both sides of the QCDSR in Eqs.~(\ref{eq:fenomvnoff}), (\ref{eq:fenomdsoff}) and (\ref{eq:piladodaqcdgeral}), which involves the following transformation of variables: $P^2 = - p^2 \to M^2$ and $P'^2 = -p'^2 \to M'^2$, where $M$ and $M'$ are the Borel masses. After that, we can equate both sides of the QCDSR invoking the quark-hadron duality. Because of the double Borel transform, quark condensate contributions to vertices $V_n D_s D_s$  are negleted when the lighter meson is off-shell. This has been taken into account when writing Eq.~(\ref{eq:contribcond}).

The second step is to eliminate the $h.r.$ terms appearing in the phenomenological side in Eqs.~(\ref{eq:fenomvnoff}) and  (\ref{eq:fenomdsoff}). This is achieved by the introduction of the continuum threshold parameters $s_0$ and $u_0$. These parameters satisfies the following relations: $m_i^2 < s_0 < {m'}_i^2$ and $m_o^2 < u_0 < {m'}_o^2$,  where  $m_i$ and $m_o$ are the masses of the incoming and outcoming mesons respectively and $m'$ is the mass of the first excited state of these mesons. Taking advantage of the quark-hadron duality, the $h.r.$ terms can be ride away from the QCDSR.

After performing these steps, we can match the phenomenological and QCD sides, obtaining the QCDSR expressions for the form factors. For example, in the case of the $p_\mu$ structure for the $V_n$ meson off-shell and the $p'_\mu$ structure for the $D_s$ meson off-shell, the obtained form factors are:
\begin{align}
\label{eq:rsqcdvn}
g_{V_n D_s D_s}^{(V_n)}(Q^2) = \frac{-\frac{3}{8\pi^2} \int^{s_0}_{s_{inf}} \int^{u_0}_{u_{inf}} \frac{1}{\sqrt{\lambda}}F_p^{(V_n)} e^{-\frac{s}{M^2}}e^{-\frac{u}{M'^2}} ds du - (n-1)m_c\langle \bar{s}s\rangle e^{-\frac{m_c^2}{M^2}} e^{-\frac{m_c^2}{M'^2}}}{\frac{(-1)^n f_{D_s}^2 f_{V_n} m^4_{D_s} m_{V_n}}{(m_c + m_s)^2(Q^2+m_{V_n}^2)}e^{-m_{D_s}^2/M^2}e^{-m_{D_s}^2/M'^2}}\\
\label{eq:rsqcdds}
g_{V_n D_s D_s}^{(D_s)}(Q^2) = \frac{-\frac{3}{8\pi^2} \int^{s_0}_{s_{inf}} \int^{u_0}_{u_{inf}} \frac{1}{\sqrt{\lambda}}F_{p'}^{(D_s)} e^{-\frac{s}{M^2}}e^{-\frac{u}{M'^2}} ds du + (2-n)m_s\langle\bar{s}s\rangle e^{-\frac{m_s^2}{M^2}} e^{-\frac{m_c^2}{M'^2}}}{\frac{(-1)^n 2 f_{D_s}^2 f_{V_n} m^4_{D_s} m_{V_n}}{(m_c + m_s)^2(Q^2+m_{D_s}^2)} e^{-m_{V_n}^2/M^2}e^{-m_{D_s}^2/M'^2}}
\end{align}
remembering that $n=1$ for $V_1=\phi$ and $n=2$ for $V_2=J/\psi$.

\section{Results and Discussion}
Eqs.~(\ref{eq:rsqcdvn}) and (\ref{eq:rsqcdds}) show the two different form factors for each vertex. In order to minimize the uncertainties when  extrapolating the QCDSR results, it is imposed that both form factors lead to the same coupling constant \cite{Bracco:2001pqq}. 
This condition was used to reduce the errors of finding the Borel masses and the continuum thresholds, which should gives a good plateau in order to obtain a physical results. Form factors should not depend on these quantities.  A good plateau is the result of a sum rule which gives a good stability. The plateau is often called ``window of stability".

Besides the values of the Borel masses and the continuum  thresholds, we need to know the values of the hadron masses and decay constants. Their values are given in Table \ref{tab:param}.
\begin{table}[ht]
\caption{\label{tab:param}Parameters used in this work, taken from Ref.~\cite{Nakamura:2010zzi}.}
\begin{ruledtabular}
\begin{tabular*}{10cm}{@{\extracolsep{\fill}} lccccc}
  & $s$ & $c$ & $\phi$ & $D_s$ & $J/\psi$ \\
\hline
  $m$ (GeV) & 0.101 & 1.27 & 1.019 & 1.96847 & 3.096916\\
  $f$ (MeV)&-&-& 229  & 257 & 416\\
\end{tabular*}
\end{ruledtabular}
\end{table}

The Borel masses, which respect the relation $M'^2 = \frac{m_o^2}{m_i^2}M^2$, can assume any value within the Borel window. This window is obtaining after imposing that the pole contribution must be bigger that the continuum contribution by $50 \%$ and the quark condensate term contribute only with 30 \% of the  perturbative term. We have worked with the mean values of the form factors within a Borel window. The procedure is briefly sketched as follows: the mean value of the form factor is calculated in the Borel window for each value of $Q^2$ used.  The standard deviation is used to automatize the analysis of the stability of the form factor with respect to the Borel masses and continuum threshold parameters. In this way, it is guarantee a good stability in the Borel window and in the whole $Q^2$ interval.

\begin{figure}[ht]
  \includegraphics[width=270px]{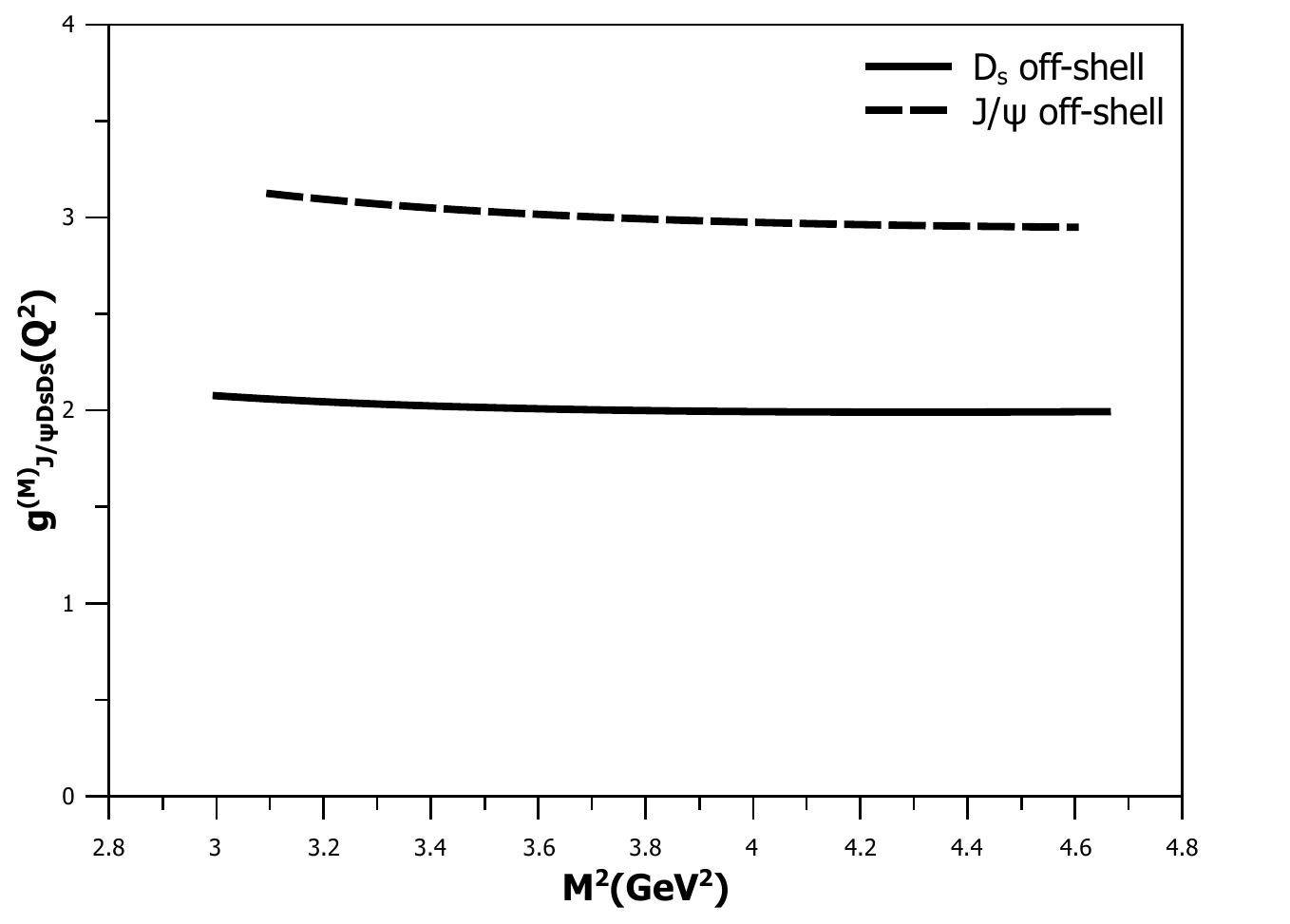}
  \caption{\label{fig:estabilidade}Form factor stability for the $J/\psi D_s D_s$ vertex at $Q^2 = 1$ GeV$^2$.}
\end{figure}

$s_0$ and $u_0$ are defining as $s_0 = (m_i + \Delta_i)^2$ and $u_0 = (m_o + \Delta_o)^2$, where  the quantities $\Delta_i$ and $\Delta_o$ have been obtained imposing the most stable sum rule.  In order to include the pole and to exclude the $h.r.$ contributions for the cases of $\phi$, $D_s$, and $J/\psi$ mesons off-shell, the values for $ \Delta_{J/\psi}$, $ \Delta_\phi$, and $\Delta_{D_s}$ cannot be far from the experimental value of the distance between the pole and the first excited state \cite{Nakamura:2010zzi,Badalian:2011tb}. Our analysis has found that the best values are $\Delta_\phi = 0.6$ GeV, $\Delta_{D_s} = 0.6$ GeV and $\Delta_{J/\psi} = 0.5$ GeV, which leads to a remarkably stable Borel windows, as can be seen in Fig.~\ref{fig:estabilidade} for the $J/\psi D_sD_s$ case.

About the choice of the Dirac structures used in the calculations, it is worthy to say that in principle, and if the whole OPE series could be summed up, both structures, $p_\mu$ and $p'_\mu$ in Eq.~(\ref{dspec}) would lead to a valid sum rule. In a real calculation however, the OPE series has to be truncated at some order, and approximations are necessary to deal with the $h.r.$ terms. Thus it is not always possible to use both structures any more. Regarding the structure $p_\mu$ for $D_s$ off-shell, it  does not lead to a coupling constant compatible with the $\phi$ off-shell case in the vertex $\phi D_s D_s$. For the others three form factors, namely $g_{J/\psi D_s D_s}^{(J/\psi)}(Q^2)$,
$g_{J/\psi D_s D_s}^{(D_s)}(Q^2)$ and $g_{\phi D_s D_s}^{(\phi)}(Q^2)$, the structure $p_\mu$ leads to results very close to the obtained using the $p'_\mu$ structure. This was taken into account when calculating the uncertainties in this work.

\subsection{Results for the $J/\psi D_s D_s$ and $\phi D_s D_s$ vertices}
In Table~\ref{tab:results} we present the $Q^2$ and Borel windows found for each form factor $g^{(M)}_{V_nD_sD_s}(Q^2)$, together with its parametrization and the respective coupling constant $g_{V_nD_sD_s}$ with its associated error $\sigma$, calculated using the method explained in subsection B. 
\begin{table}[ht]
\caption{\label{tab:results}Parametrization of the form factors and numerical results for the coupling constant of this work. The calculation of $\sigma$ is explained in subsection B.}
\begin{ruledtabular}
\begin{tabular*}{\linewidth}{@{\extracolsep{\fill}} ccccc}  
                      & \multicolumn{4}{c}{Vertex} \\ 
\cline{2-5}  
                      & \multicolumn{2}{c}{$J/\psi D_s D_s$} & \multicolumn{2}{c}{$\phi D_s D_s$} \\
Quantity                      & $J/\psi$ off-shell & $D_s$ off-shell & $\phi$ off-shell & $D_s$ off-shell \\ 
\hline
$Q^2$ (GeV$^2$)            & [0.5, 2.0]    & [1.0, 5.0]    & [1.0, 2.5]    & [0.1, 1.6]\\
$M^2$ (GeV$^2$)            & [3.1, 4.6]    & [3.0, 4.7]    & [1.2, 2.0]    & [0.6, 1.3]\\
$g^{(M)}_{V_nD_sD_s}(Q^2)$ & $A e^{-Q^2/B}$ & $A e^{-Q^2/B}$ & $A e^{-Q^2/B}$ & $\frac{A}{B+Q^2}$\\
$A$                          & 3.21          & 2.503         & 1.091         & 16.25 GeV$^2$\\
$B$ (GeV$^{2}$)             & 15.52         & 4.425         & 1.889         & 13.3 \\
${g^{(M)}_{V_nD_sD_s}}\pm\sigma$     & $5.96^{+0.97}_{-0.91}$          & $6.01^{+0.52}_{-0.43}$     &     $1.89^{+0.18}_{-0.14}$ & $1.73^{+0.09}_{-0.09}$\\
\end{tabular*}
\end{ruledtabular}
\end{table}

\begin{figure}[ht]
  \includegraphics[width=231px]{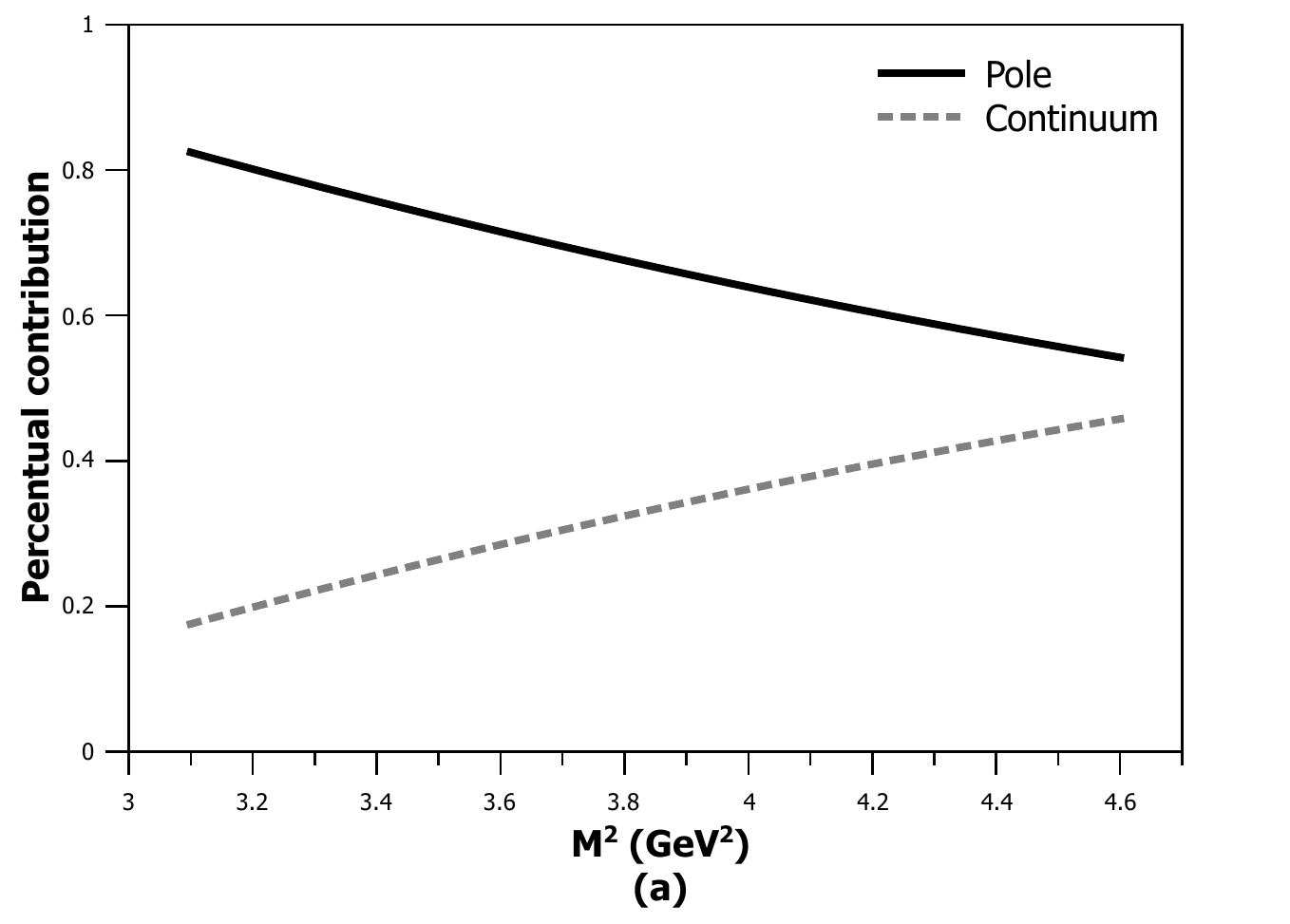}
  \includegraphics[width=231px]{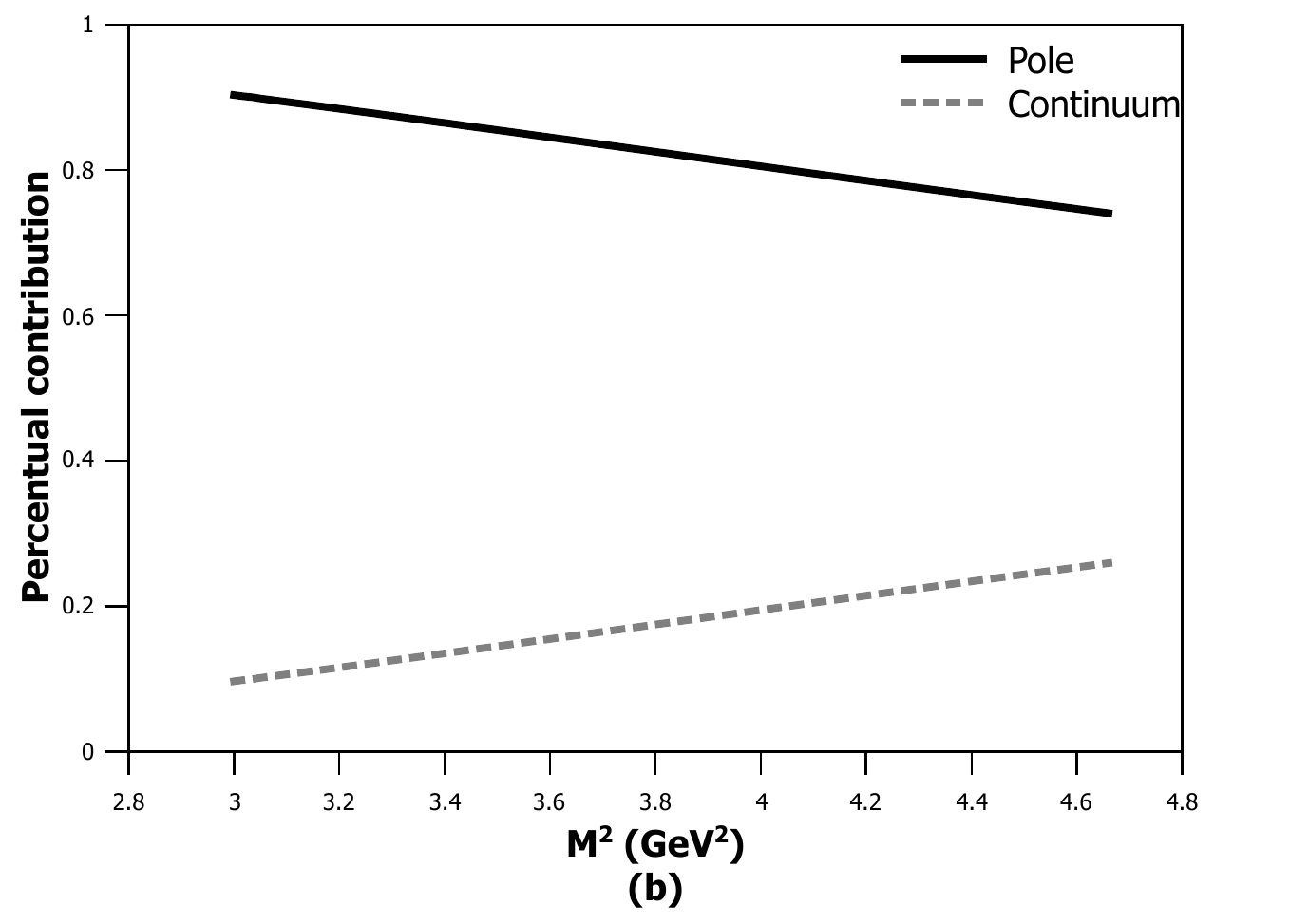}
  \caption{\label{fig:jpsidsdspolocont}Pole and continuum contributions for the $J/\psi$ off-shell (panel \textbf{(a)}) and for $D_s$ off-shell (panel \textbf{(b)}), both for the $J/\psi D_s D_s$ vertex at $Q^2 = 1$ GeV$^2$.}
  \end{figure}

The Borel windows presented in Table~\ref{tab:results} ($M^2$ row) satisfies the already mentioned conditions regarding the pole and continuum contributions, as can be seen in Fig.~\ref{fig:jpsidsdspolocont} for the $J/\psi D_s D_s$ vertex. For the $\phi D_sD_s$ case we obtained similar results. These Borel windows also satisfies the dominance relations between the perturbative over the condensate contributions by at least 70\% of the total contribution for both vertices.

\begin{figure}[ht]
  \includegraphics[width=231px]{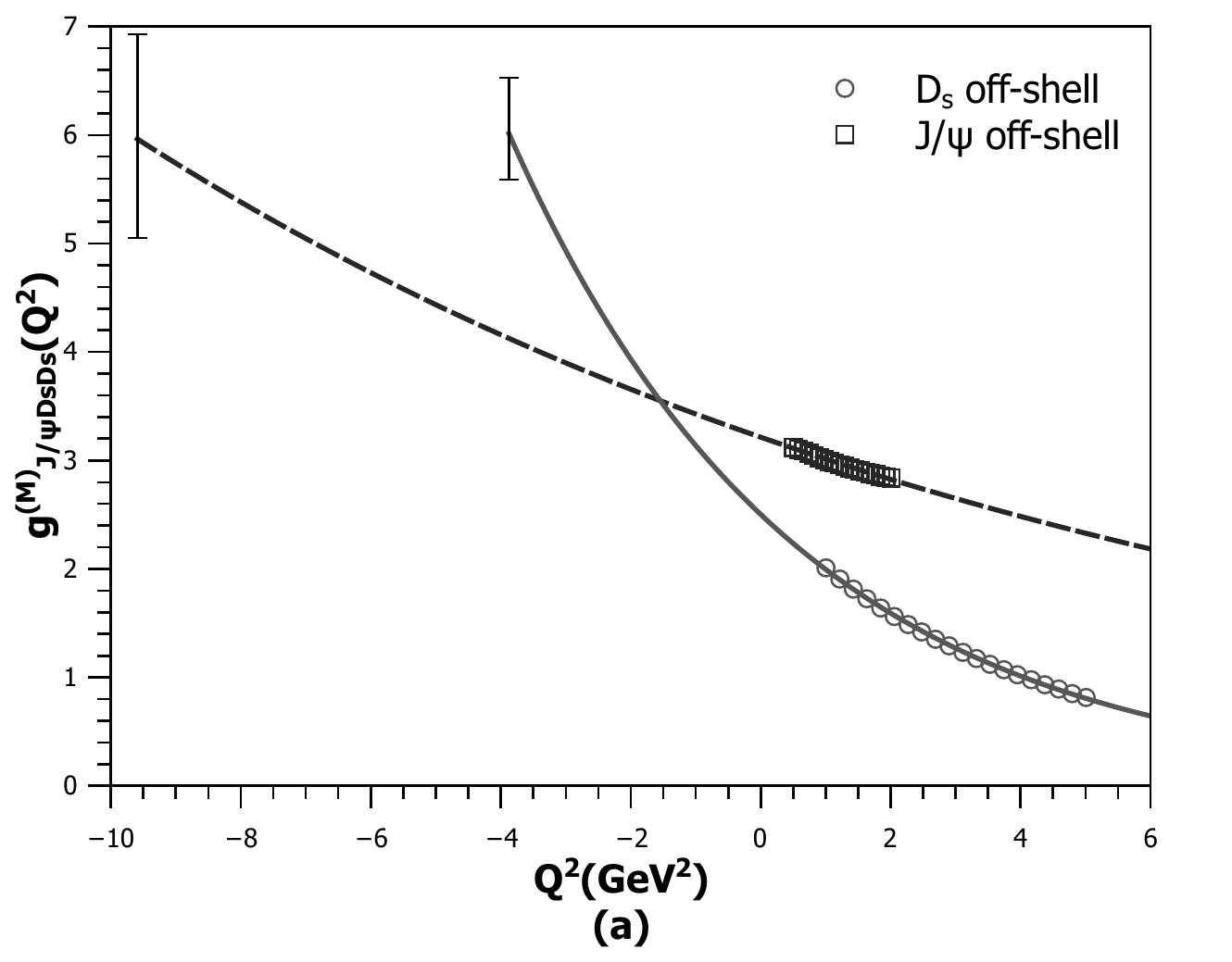}
	\includegraphics[width=231px]{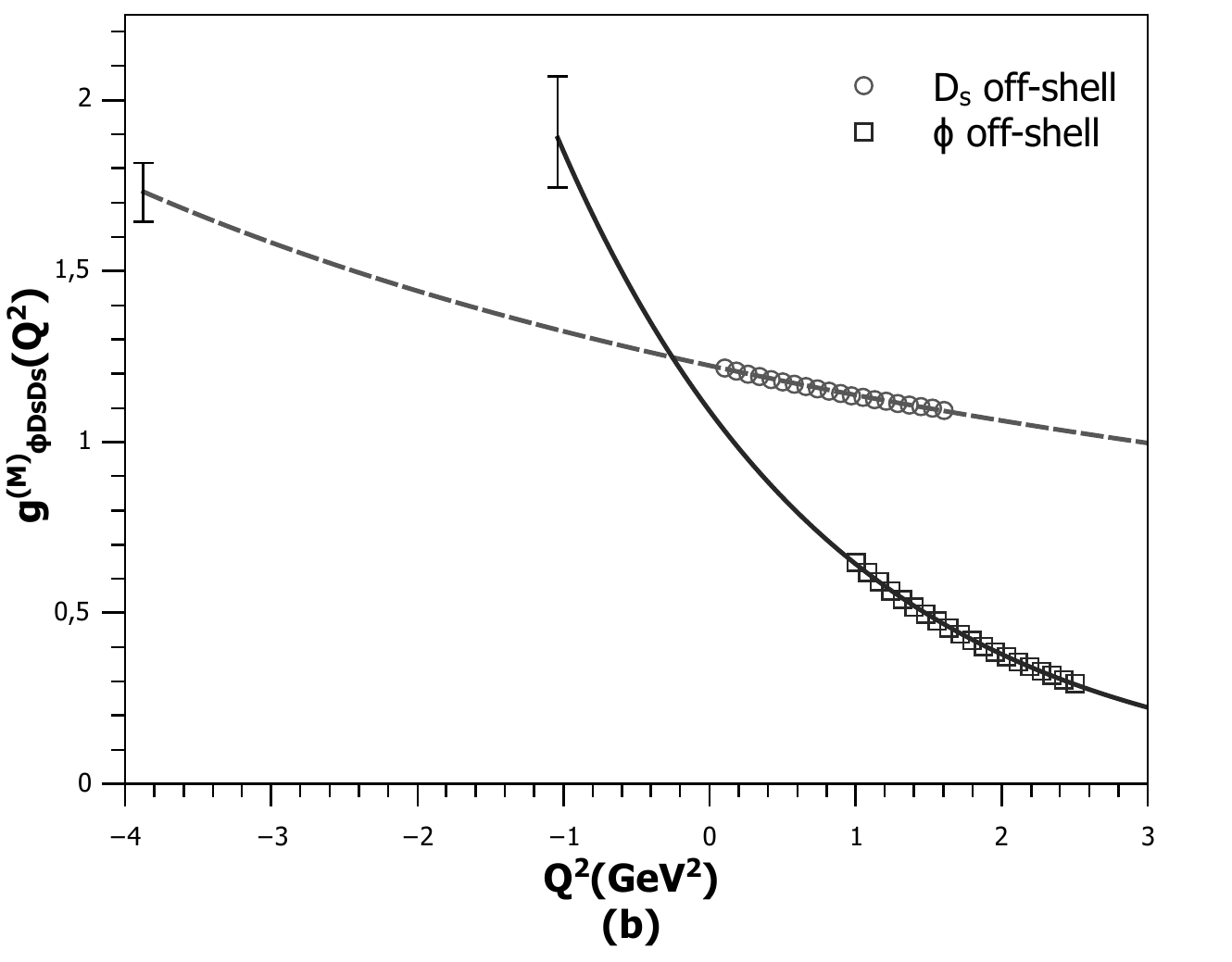}
  \caption{\label{fig:formfactors}Form factors for the $J/\psi D_s D_s$ vertex (panel \textbf{(a)}) and for the $\phi D_s D_s$ vertex (panel \textbf{(b)}).}
  \end{figure}

The form factors obtained for the $J/\psi D_s D_s$ vertex, with $J/\psi$ and $D_s$ off-shell, were both well adjusted by exponential curves (Fig.~\ref{fig:formfactors}(a)). For the vertex $\phi D_s D_s$, the form factor for the case $D_s$ off-shell was fitted by a monopolar curve while for $\phi$ off-shell it was adjusted by an exponential curve (Fig.~\ref{fig:formfactors}(b)). 

\subsection{Error analysis}
The vertex $J/\psi D_s D_s$ has similarities with the vertex $\phi D_s D_s$. However, they are physically different in such a way that it is expected different values for their coupling constants. Nevertheless, the two coupling constants of the same vertex should have the same value, independently of which meson is off-shell. The coupling constants presented in Table~\ref{tab:results} have different values among them even when we look for the two coupling constants of the same vertex. But when the errors are taken into account, we can see that they are compatible, as suggested by Fig.~\ref{fig:formfactors}, where the error bars stands for the uncertainties of the coupling constants in Table~\ref{tab:results}.

In works \cite{Rodrigues:2010ed,CerqueiraJr2012130}, we estimated the errors by studying the behavior of the coupling constant when each one of the parameters involved in the calculation were varied individually between their upper and lower limits. Experimental parameters have their own errors (showed in Table~\ref{tab:errors}), while for the QCDSR parameters, we varied the thresholds $\Delta_s$ and $\Delta_u$ in $\pm 0.1$ GeV, the momentum $Q^2$ in $\pm 20\%$ and for error due to the Borel mass $M^2$ we used the standard deviation of the average value of the form factor within the Borel window. Then, we calculated the mean value of all these contributions and their standard deviations. The same procedure was adopted in this work in other to estimate the error bars. In Table~\ref{tab:errors}, we can see how the variation of each parameter affects the final value of the coupling constants.
\begin{table}[ht]
\caption{\label{tab:errors}Percentage deviation of the coupling constants related with each parameter.}
\begin{ruledtabular}
\begin{tabular*}{\linewidth}{@{\extracolsep{\fill}} ccccc}
                   &    \multicolumn{4}{c} {Deviation {(\%)}} \\
\cline{2-5}
 Parameter                    & $\Delta g_{J/\psi D_s D_s}^{(J/\psi)}$ & $\Delta g_{J/\psi D_s D_s}^{(D_s)}$ & $\Delta g_{\phi D_s D_s}^{(\phi)}$ & $\Delta g_{\phi D_s D_s}^{(D_s)}$ \\ 
\hline
$f_{D_s} = 257.5 \pm 6.1$(MeV)     & 3.88  & 3.87  & 3.88 & 3.86 \\
$f_{J/\psi} = 416 \pm 6 $(MeV)     & 1.18  & 1.18  & - & - \\
$f_{\phi} = 229 \pm 3$(MeV)        & -     &   -   & 1.07 & 1.07\\
$m_c = 1.27^{+0.07}_{-0.09}$ (GeV) & 14.7  & 18.9  & 15.4 & 11.1\\
$m_s = 101^{+29}_{-21}$ (MeV)      & 1.25  & 3.10  & 5.43 & 3.37\\
$M^2$ (GeV$^2$)\footnotemark[1]                  & 6.46  & 7.70  & 6.56 & 0.72\\
$\Delta_s \pm 0.1$(GeV)            & 32.5  & 2.70  & 9.56 & 3.09\\
$\Delta_u \pm 0.1$(GeV)            & 26.7  & 2.08  & 12.0 & 0.46\\
$Q^2 \pm 20\%$ (GeV$^2$)\footnotemark[1]           & 3.38  & 2.34  & 2.91 & 1.54\\
$\langle s\bar{s}\rangle = (0.8 \pm 0.2)(0.23 \pm 0.03)^3$ (GeV$^3$) & 7.70 & - & - & 5.00\\
Fitting parameters  ($A$ and $B$)      & 0.94  & 0.11  & 0.15 & 3.12\\
\end{tabular*}
\footnotetext[1]{The intervals for these quantities are those of Table~\ref{tab:results}.}
\end{ruledtabular}
\end{table}

It follows from Table~\ref{tab:errors} that the variation of most of the parameters (namely, the decay constants, the Borel  and the strange quark masses for example) have little impact on the value of the coupling constants. We have found however that the error in the charm mass is the one with the biggest propagation over the final value of the coupling constants. The fact that the sensitivity related to the Borel mass is small  was already expected, as we obtained a very stable Borel windows for both vertices (Fig.~\ref{fig:estabilidade}). 

All the results showed until this point are for the structure $p_\mu$ in the case $V_n$ off-shell and for the structure $p'_\mu$ in the case $D_s$ off-shell. As said earlier, we  can also work with  other (good) structures in order to obtain the final value of $g_{J/\psi D_s D_s}$ and $g_{\phi D_s D_s}$ and their uncertainties. In the case of $J/\psi D_s D_s$, we used the coupling constant obtained from the study of the structure $p_\mu$ in the $D_s$ off-shell case, which value is:
\begin{align}
g_{J/\psi D_s D_s}^{(D_s)} = 6.56^{+0.57}_{-0.47} 
\label{eq:gPDsoff}
\end{align}
and also the structure $p'_\mu$ for the $J/\psi$ off-shell case, which value is:
\begin{align}
g_{J/\psi D_s D_s}^{(J/\psi)} = 6.10^{+1.07}_{-1.01} 
\label{eq:gPlJpsioff}
\end{align}

For the $\phi$ off-shell case in the vertex $\phi D_s D_s$ we used also the result obtained from the $p'_\mu$ structure, which reads:
\begin{align}
g_{\phi D_s D_s}^{(\phi)} = 1.90^{+0.17}_{-0.13}
\label{eq:gPlPhioff}
\end{align}

\section{Conclusion}
Using the results presented in Table~\ref{tab:results} and in Eqs.~(\ref{eq:gPDsoff}), (\ref{eq:gPlJpsioff}) and (\ref{eq:gPlPhioff}), we taken the mean value of the calculated coupling constants, obtaining the following final results for the coupling constants of the two vertices studied in this work:
\begin{align}
g_{J/\psi D_s D_s} = 6.20^{+0.97}_{-1.15}
\label{eq:ctejpsidsdsd}\\
g_{\phi D_s D_s} = 1.85^{+0.22}_{-0.23}
\label{eq:ctephidsdsd}
\end{align}
We call the attention again to the fact that these results were obtained from a ``good" QCDSR, which means a good pole/continuum dominance, a perturbative contribution greater than the condensate contribution and a very stable Borel window for the whole $Q^2$ interval studied. Also, the coupling constants with the $D_s$ meson off-shell were compatible with the coupling constants obtained for the vector meson off-shell case in both vertices. Regarding the form factors, we observe the same behavior found in our previous works: the form factor is harder when the heavier meson is off-shell. The errors are of the order of $10\%$-$20\%$, what are the expected values coming from previous QCDSR works. 

The results of Eqs.~(\ref{eq:ctejpsidsdsd}) and (\ref{eq:ctephidsdsd}) can be compared with the ones coming from other calculations. For example, the result for the $g_{\phi D_s D_s}$ of this work can be directly compared with the result coming from the Light Cone QCDSR (LCQCDSR) formalism, as it is shown in Table~\ref{tab:otherworks}, concluding that they are in agreement.
\begin{table}[h!]
\caption{\label{tab:otherworks}Meson coupling constants obtained from other QCDSR calculations.}
\begin{ruledtabular}
\begin{tabular*}{13.0cm}{@{\extracolsep{\fill}} ccc}    
QCDSR \cite{Matheus:2005yu} & QCDSR \cite{Bracco:2004rx} & LCQCDSR \cite{Wang:2007zm}\\ \hline
$g_{J/\psi D D} = 5.8 \pm 0.8$ & $g_{J\psi D^* D^*} = 6.2 \pm 0.9$ & $g_{\phi D_s D_s} = 1.45 \pm 0.34$\\
\end{tabular*}
\end{ruledtabular}
\end{table}
We can use also previous QCDSR results for other vertices to compare with our $g_{J/\psi D_s D_s}$ value. For example, invoking the (broken) SU(3) symmetry \cite{Liu:2006dq}, it is expected the relation $g_{J/\psi D_s D_s} = g_{J/\psi D D}$.  Using the value for $g_{J/\psi D D}$ from Table~\ref{tab:otherworks}, we can see that this relation is sustained. Still inside the SU(3) scheme, the relation $g_{J/\psi D_s D_s} = g_{J/\psi D^* D^*}$ should be valid \cite{Bracco:2011pg,Liu:2006dq}. Comparing our result for $g_{J/\psi D_s D_s}$ with the one for $g_{J/\psi D^* D^*}$ from Table~\ref{tab:otherworks}, we obtain again compatible results. The $g_{J/\psi D_s D_s} = g_{J/\psi D D}$ SU(3) relation is violated by the order of $8\%$, the same value found between $g_{J\psi D^* D^*}$ and $g_{J/\psi D D}$ \cite{Bracco:2011pg}, which means that, in spite of the huge mass difference between the involved mesons, SU(3) is a reasonable symmetry for this vertices when using QCDSR.

\acknowledgments
This work has been partially supported by CNPq and CAPES.

\end{document}